\documentclass{article}
\usepackage[table]{xcolor}
\usepackage{spconf,amsmath,graphicx,float,amssymb, algorithm, algorithmicx}
\usepackage{array}
\usepackage{comment}
\usepackage{booktabs}
\usepackage[noend]{algpseudocode}
\usepackage[T1]{fontenc}
\usepackage{bbm}
\usepackage{mathrsfs}
\usepackage{arydshln}
\usepackage{hyperref}
\usepackage[nolist]{acronym}

\begin{acronym}[nolist]
    \acro{CXR}{Chest X-Ray}
\end{acronym}

\definecolor{greencolor}{HTML}{bdeb98}
\definecolor{redcolor}{HTML}{ff9896}
\definecolor{bluecolor}{HTML}{c1e4f7}
\definecolor{orangecolor}{HTML}{ffbb78}
\definecolor{pinkcolor}{HTML}{fccde5}

\title{Fake-X-RayVerse: Detecting Subtle Anomalies and Fine-Grained Deepfakes in Medical Radiographs}

\name{\begin{tabular}{c}
  Devansh Arora$^{*1}$, Nitin Kumar$^{*1}$, Anubhav Kataria$^{2}$, Nishant Rana$^{3}$, \\
  Deepander Singh Rathore$^{3}$, Sukrit Gupta$^{1}$, Abhinav Dhall$^{2}$
\end{tabular}}
\address{$^{1}$Indian Institute of Technology Ropar, Rupnagar, India \\
         $^{2}$Monash University, Melbourne, Australia \\
         $^{3}$Homi Bhabha Cancer Hospital \& Research Centre, New Chandigarh, Punjab, India}

\begin{document}
%
\maketitle
{\renewcommand{\thefootnote}{*}\footnotetext{Equal contribution.}}
\begin{abstract}
Multimodal medical deepfakes that jointly generate or manipulate chest X-Rays (CXRs) and their associated radiology reports pose a distinct challenge: individually plausible modalities may still be clinically inconsistent with each other or with the intended pathology. We present a systematic evaluation of multimodal CXR deepfakes using both AI model-based and human evaluation. Using MIMIC-CXR, we evaluate three generation pipelines, CXR-IRGen, ProgEmu, and a Qwen-based image-editing and report-generation pipeline, across four dimensions: image–report coherence, pathology label fidelity, image distributional fidelity, and report quality. 
CXR-IRGen performs strongest on most automated metrics, while ProgEmu achieves higher image distributional fidelity. In contrast, human raters find Qwen-generated samples hardest to distinguish from real images, revealing a gap between automated metrics and human detectability. All three generation methods also exhibit substantial degradation in image–report coherence compared with real data. These findings reveal a gap between automated measures of generative fidelity and human detectability, and highlight the need for multimodal, clinically grounded benchmarks for evaluating medical deepfakes. GitHub: \href{https://github.com/AI4Med-Lab/CXR-Deepfake-Survey}{https://github.com/AI4Med-Lab/CXR-Deepfake-Survey}

\end{abstract}
\section{Introduction}
\label{sec:intro}

Generative models are increasingly capable of synthesizing and editing medical images, including \ac{CXR}, with sufficient visual fidelity to challenge conventional approaches for distinguishing real from generated radiographs. Recent work has demonstrated both direct generation of \ac{CXR} and targeted manipulation of pathological findings~\cite{shentu2024cxrirgen,ma2025progemu,tordjman2026deepfake,chen2026medforge}. While these capabilities have legitimate applications in medical education and research, they also create opportunities for fabricated or manipulated radiological evidence to appear clinically authentic~\cite{tordjman2026deepfake,chen2026medforge,li2025medforensics}.

The challenge becomes substantially more complex when a generated image is accompanied by a radiology report. A medical deepfake is not necessarily convincing simply because its individual modalities appear realistic: the report must describe the image correctly, and both modalities should reflect the pathology that was intended during generation. For example, an X-Ray may plausibly depict pulmonary edema while its accompanying report describes a normal study, or an apparently coherent image–report pair may fail to contain the pathology specified in the generation prompt. Thus, evaluating medical deepfakes requires assessing not only the realism of individual modalities, but also their pathological fidelity and cross-modal consistency.

Existing medical deepfake studies have primarily focused on image-level generation, manipulation, and detection~\cite{chen2026medforge,li2025medforensics}. Similarly, recent multimodal generation approaches such as CXR-IRGen~\cite{shentu2024cxrirgen} and ProgEmu~\cite{ma2025progemu} demonstrate the ability to generate \ac{CXR}s together with textual reports, but a systematic comparison of their image quality, report quality, pathology fidelity, and image–report coherence remains limited. Moreover, automated measures of generative quality may not necessarily reflect how readily humans can identify medical deepfakes. Understanding this relationship is particularly important for clinical applications, where subtle inconsistencies may be more consequential than generic image-level artifacts.

In this work, we conduct a systematic evaluation of methods for deepfake generation \ac{CXR}s and the text reports. Using MIMIC-CXR \cite{johnson2019mimic}, we evaluate CXR-IRGen~\cite{shentu2024cxrirgen}, ProgEmu~\cite{ma2025progemu}, and a Qwen-based image-editing and report-generation pipeline across four dimensions: (i) image–report coherence, (ii) fidelity to the intended pathology, (iii) image and text distributional fidelity, and (iv) textual quality. We complement automated evaluation with a blinded human study involving a clinician and a radiology trainee, assessing real-versus-fake discrimination, image–report matching, and pathology identification. We further investigate multimodal deepfake detection using a medical vision–language model. 

Our results show substantial degradation in image–report coherence across all generation methods relative to real data. Our findings highlight a measurable gap between automated assessments of generative fidelity and human detectability and motivate multimodal, clinically grounded evaluation of medical deepfakes.

\begin{figure*}[h]
    \centering
    \includegraphics[width=0.85\linewidth]{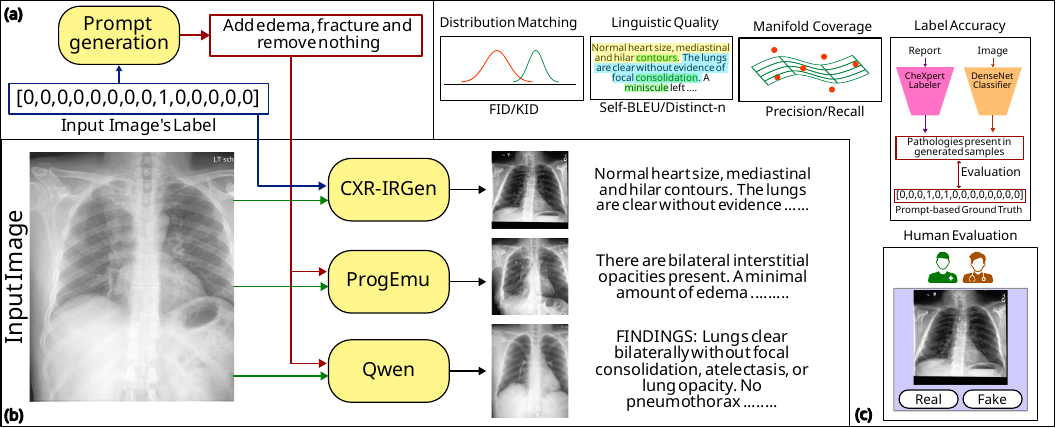}
    \caption{\textbf{Complete deepfake generation pipeline and evaluation protocols.} (a) First, an editing prompt is generated by an LLM using the input image's label. (b) The generated prompt and the input image label guide the models for deepfake generation. (c) The generated deepfakes are evaluated using distribution matching methods, manifold coverage evaluation, linguistic quality, and label accuracy tests followed by an expert evaluation study.\protect\footnotemark[1]}
    \label{fig:pipeline}
\end{figure*}

\section{Methodology}
\label{sec:method}
\subsection{Deepfake Generation Pipelines}
\label{sec:paradigms}

\noindent\textbf{CXR-IRGen.} CXR-IRGen \cite{shentu2024cxrirgen} combines a diffusion-based image generation module with a BART-based report generation module. It takes a real source image together with its ground-truth pathology label(s) as input, and generates a new synthetic image-report pair conditioned to preserve the same diagnostic label(s).

\noindent\textbf{ProgEmu.} ProgEmu \cite{ma2025progemu} is an autoregressive multimodal model that jointly generates a counterfactual image and its accompanying interpretive report. It takes as input a real source image and an editing prompt describing a pathology to inject or remove.

\noindent\textbf{Qwen pipeline.} Our Qwen-based pipeline uses Qwen-Image-Editing-2511~\cite{wu2025qwen} for image editing and Qwen3-VL-30B-A3B-Thinking~\cite{bai2025qwen3} for report generation. Qwen-Image-Editing-2511 takes the source image and editing prompt to generate the counterfactual deepfake. Qwen3-VL-30B-A3B-Thinking then generates the accompanying report from the edited image.


Editing instructions for the prompt-driven pipelines are generated by Qwen-2.5-Instruct using the source image's original ground-truth pathology labels as context. The output instruction is constrained to a fixed format: [ADD $\langle$class(es) or `nothing'$\rangle$ and REMOVE $\langle$class(es) or `nothing'$\rangle$]. Pathology labels are derived from this instruction to serve as ground truths for deepfake evaluations.
\subsection{Deepfake Quality Evaluation}
We evaluate the generated deepfake images and reports through four-fold analysis:

\noindent\textbf{Image-Report Coherence}: We embed images and reports into a shared latent space using MedSigLIP~\cite{sellergren2025medgemma}, a SigLIP-style dual encoder adapted for the medical domain, and evaluate cross-modal coherence via what we term \emph{Cross-Modal Optimal Assignment} (CMOA). For a round of size $n$, we sample $n$ image-report pairs from a single method with non-overlapping pathology classes, embed them and compute the cosine similarity matrix $S \in \mathbbm{R}^{n \times n}$. We solve $\hat{\sigma} = \arg\max_{\sigma \in \Pi_n} \sum_{i=1}^{n} S_{i, \sigma(i)}$ via the Hungarian algorithm~\cite{kuhn1955hungarian}, where $\Pi_n$ is the set of permutations of $\{1, \ldots, n\}$, giving the optimal assignment. 

\noindent\textbf{Label Fidelity}: We assess agreement between the diagnostic content of generated images/reports and their intended (ground-truth or prompted) label set against the predicted set, via accuracy and macro-averaged precision, recall, and F1. We classify the images using  DenseNet121-Res224~\cite{cohen2021torchxrayvision}. We label the generated text with the CheXpert-Labeler~\cite{irvin2019chexpert}, which was originally used to label to MIMIC-CXR dataset.

\noindent\textbf{Image and Text Realism}: We assess distributional similarity between real and generated images by embedding both sets with a fixed pretrained image encoder resulting in embedding distributions $P_r$ (real) and $P_g$ (generated). We compute the Fr\'echet Inception Distance (FID) between Gaussian approximations fit to $\{f(\boldsymbol{x})\}_{\boldsymbol{x} \sim P_r}$ and $\{f(\boldsymbol{x})\}_{\boldsymbol{x} \sim P_g}$. We further compute the Kernel Inception Distance (KID) between $P_r$ and $P_g$ in embedding space. We additionally report precision and recall by forming $k$-nearest-neighbor manifold approximations of $\{f(\boldsymbol{x})\}_{\boldsymbol{x} \sim P_r}$ and $\{f(\boldsymbol{x})\}_{\boldsymbol{x} \sim P_g}$ from hyperspheres centered at each embedded sample. Together, these metrics are standard for assessing the realism of deepfakes~\cite{pei2024deepfake}.

\noindent\textbf{Linguistic Quality}: We further evaluate generated reports for diversity and intrinsic linguistic quality. We compute Self-BLEU score by treating each report in turn as a hypothesis sample scored against the remaining reports as references. Further we  compute Distinct-$n$ for $n \in \{1, 2\}$ as the ratio of unique to total $n$-grams across the corpus to capture lexical diversity. For intrinsic linguistic quality, we report the Flesch Reading Ease score together with a grammar error rate and a repetition ratio. 

\noindent\textbf{Expert evaluation study}: To further assess the quality of deepfake generation, we conduct a blinded human study with two independent raters, a trained clinician and a postgraduate radiology trainee, conducted on a self-made website, evaluating samples without access to method labels. The study spans three tasks: (i) $5 \times 5$ image-report matching, (ii) pathology identification against the MIMIC-CXR label set, and (iii) real-versus-fake discrimination on images and reports separately.
\subsection{Deepfake Detection Pipeline}
\label{sec:detector}

We train a binary (real vs.\ fake) classification head on a dataset pooling synthetic samples from all three methods with real MIMIC-CXR images, by finetuning pretrained backbone image and text encoders, $f_{\mathrm{img}}: \mathcal{X} \rightarrow \mathbb{R}^{z}$ and $f_{\mathrm{txt}}: \mathcal{R} \rightarrow \mathbb{R}^{z}$, mapping into a shared embedding space. For an image--report pair $(\boldsymbol{X}, \boldsymbol{r}) \in \mathcal{X} \times \mathcal{R}$, we obtain embeddings $\boldsymbol{u} = f_{\mathrm{img}}(\boldsymbol{X})$ and $\boldsymbol{v} = f_{\mathrm{txt}}(\boldsymbol{r})$, with $\boldsymbol{u}, \boldsymbol{v} \in \mathbb{R}^{z}$, and train a multimodal detection head $g_\theta^{\mathrm{multi}}: \mathbb{R}^{2z} \rightarrow [0,1]$ on the concatenated embedding $\boldsymbol{z} = [\boldsymbol{u} \,;\, \boldsymbol{v}] \in \mathbb{R}^{2z}$. 

We evaluate this multimodal detector against two unimodal baselines to test whether joint image-report information improves detectability over either modality alone, by similarly finetuning a modality-specific encoder $h_m: \mathcal{M} \rightarrow \mathbb{R}^{z_m}$ for $m \in \{\mathrm{img}, \mathrm{txt}\}$ (with $\mathcal{M} = \mathcal{X}$ or $\mathcal{R}$ accordingly), each paired with its own head $g_{\phi_m}^{\mathrm{uni}}: \mathbb{R}^{z_m} \rightarrow [0,1]$ trained on $h_m(\cdot)$ alone. All models are trained to maximize the expected log-likelihood of the ground-truth authenticity label $y \in \{0,1\}$ using cross-entropy loss.
\section{Experiments and Results}
\label{sec:pagestyle}

\subsection{Dataset}
\label{sec:dataset}
We use the MIMIC-CXR dataset \cite{johnson2019mimic}, a large, publicly available collection of chest radiographs paired with radiology reports and CheXpert-derived pathology labels \cite{irvin2019chexpert}. MIMIC-CXR spans multiple projections, with 96,155 images labeled as posteroanterior (PA) in the official dataset metadata. We use only the PA view, since each model used in this work imposes its own constraints, and only the PA view was uniformly supported across all pipelines. Image preprocessing were left to each individual pipeline's own method, so as to evaluate each pipeline as released. As a part of this study, we release a dataset of 6000 deepfake \ac{CXR}-Radiology Report pairs generated through the described methods, which can be accessed at: \href{https://huggingface.co/datasets/AI4Med-Lab/CXR-Deepfake-Images}{https://huggingface.co/datasets/AI4Med-Lab/CXR-Deepfake-Images}. 


\subsection{Experimental Setup}
\label{sec:setup}

\footnotetext[1]{Representative images sourced from the public Indiana University CXR dataset~\cite{DemnerFushman2016} in compliance with MIMIC-CXR data use policies.}

We utilize two NVIDIA RTX PRO 6000 Blackwell GPUs for model training and inference. We used frozen pretrained weights of the models being assessed for deepfake generation. We validated the generated samples across four axes, as reported in the subsequent section. We trained our deepfake detection models on 18K image-report pairs having a ratio of 5:1 for real \& fake pairs, with metrics reported for a test set comprising 4K pairs having equally sampled instances from each of the generation methods and real samples.
\subsection{Deepfake Analysis}
The results are organized in various tables. Each table highlights the best performing method with \textbf{bold} and the second-best performing method with \underline{underline}. 

\noindent\textbf{Image-Report coherence}: Table~\ref{tab:matching} reports the matching accuracy averaged over 50 rounds, for $n \in \{3, 4, 5\}$. We also report the average score achieved by human evaluators on the $5 \times 5$ matching task.

\begin{table}[t]
\centering
\caption{Image-report matching accuracy (\%)}
\label{tab:matching}
\resizebox{\columnwidth}{!}{%
\begin{tabular}{lcccc}
\toprule
\textbf{Method} & $\mathbf{3\times3}\uparrow$ & $\mathbf{4\times4}\uparrow$ & $\mathbf{5\times5}\uparrow$ & \textbf{Evaluators}$\uparrow$ \\
\midrule
\rowcolor{redcolor} Real      & 85.00 & 80.00 & 78.40 & 45.00 \\
CXR-IRGen & \textbf{67.00} & \textbf{58.00} & \textbf{46.80} & \underline{32.50} \\
ProgEmu   & \underline{42.00} & \underline{38.50} & \underline{32.80} & 25.00 \\
Qwen      & 40.00 & 31.50 & 27.60 & \textbf{52.50} \\
\bottomrule
\end{tabular}
}
\end{table}

\noindent\textbf{Label Fidelity}: Table~\ref{tab:image-label} reports the evaluation metrics recorded for diagnostic alignment of generated images with the ground-truth. it also presents the human evaluators' average sample F1 score on the same task. Evaluation metrics for label alignment of generated reports are presented in Table~\ref{tab:text-label}.

\begin{table}[h]
\centering
\caption{Agreement between pathology labels of generated images and ground-truth/prompted labels.}
\label{tab:image-label}
\resizebox{\columnwidth}{!}{%
\begin{tabular}{lccccc}
\toprule
\textbf{Model} & \textbf{Acc.}$\uparrow$ & \textbf{Prec.}$\uparrow$ & \textbf{Rec.}$\uparrow$ & \textbf{F1}$\uparrow$ & \textbf{Eval. Acc.}$\uparrow$ \\
\midrule
\rowcolor{redcolor} Real     & 0.66 & 0.08 & 0.76 & 0.14 & 0.75\\
CXR-IRGen & 0.54 & 0.12 & \textbf{0.82} & \textbf{0.19} & \textbf{0.20}\\
ProgEmu  & \underline{0.71} & \underline{0.13} & \underline{0.56} & \underline{0.18} & \underline{0.18}\\
Qwen     & \textbf{0.72} & \textbf{0.14} & 0.23 & 0.12 & 0.13 \\
\bottomrule
\end{tabular}}
\end{table}

\begin{table}[h]
\centering
\caption{Agreement between CheXpert-Labeler classification of generated reports and ground-truth/prompted labels.}
\label{tab:text-label}
\begin{tabular}{lcccc}
\toprule
\textbf{Model} & \textbf{Acc.}$\uparrow$ & \textbf{Prec.}$\uparrow$ & \textbf{Rec.}$\uparrow$ & \textbf{F1}$\uparrow$ \\
\midrule
\rowcolor{redcolor} Real & 1.00 & 1.00 & 1.00 & 1.00 \\ 
CXR-IRGen & \textbf{0.79} & \underline{0.57} & \underline{0.61} & \textbf{0.57} \\
ProgEmu  & \underline{0.70} & \textbf{0.59} & \textbf{0.73} & \underline{0.57} \\
Qwen     & 0.54 & 0.52 & 0.55 & 0.44 \\
\bottomrule
\end{tabular}
\end{table}

\noindent\textbf{Image and Text Realism}: We utilize RAD-DINO and BiomedCLIP to extract meaningful features for distributional analysis of generated images. The results of this analysis are presented in Table~\ref{tab:image-fidelity}. Table~\ref{tab:text-fidelity} reports similar analysis for generated reports in CXR-BERT's latent space. Precision and Recall are calculated in the native space for each distribution. 


\begin{table}[h]
\centering
\caption{Image fidelity metrics. RD = RAD-DINO features; BC = BiomedCLIP features.}
\label{tab:image-fidelity}
\resizebox{\columnwidth}{!}{%
\begin{tabular}{lcccccc}
\toprule
\textbf{Model} & \textbf{FID}$_{\text{RD}}$$\downarrow$ & \textbf{KID}$_{\text{RD}}$$\downarrow$ & \textbf{FID}$_{\text{BC}}$$\downarrow$ & \textbf{KID}$_{\text{BC}}$$\downarrow$ & \textbf{Prec.}$\uparrow$ & \textbf{Rec.}$\uparrow$ \\
\midrule
\rowcolor{redcolor} Real & 0 & 0 & 0 & 0 & 1.00 & 1.00 \\
CXR-IRGen & \underline{101.82} & \textbf{0.11} & \textbf{618.87} & \textbf{327.81} & \underline{0.40} & 0.04 \\
ProgEmu & \textbf{91.68} & \underline{0.14} & 1896.15 & 1351.67 & \textbf{0.90} & \textbf{0.15} \\
Qwen & 117.05 & 0.200 & \underline{869.24} & \underline{534.44} & 0.33 & \underline{0.12} \\
\bottomrule
\end{tabular}%
}
\end{table}

\begin{table}[t]
\centering
\caption{Text fidelity metrics.}
\label{tab:text-fidelity}
\begin{tabular}{lcccc}
\toprule
\textbf{Model} & \textbf{FID}$\downarrow$ & \textbf{KID}$\downarrow$ & \textbf{Prec.}$\uparrow$ & \textbf{Rec.}$\uparrow$ \\
\midrule
\rowcolor{redcolor} Real      & 0      & 0      & 1.00 & 1.00 \\
CXR-IRGen & \textbf{0.18} & \textbf{0.00} & \underline{0.88} & \textbf{0.81} \\
ProgEmu   & \underline{0.53} & \underline{0.01} & 0.86 & 0.50 \\
Qwen      & 0.95 & 0.01 & \textbf{0.89} & \underline{0.67} \\
\bottomrule
\end{tabular}
\end{table}

\noindent\textbf{Linguistic Quality}: Table~\ref{tab:text-quality} reports the linguistic quality metrics for the real and generated reports.

\begin{table}[!h]
\centering
\caption{Diversity and intrinsic linguistic quality metrics for generated reports.}
\label{tab:text-quality}
\resizebox{\columnwidth}{!}{%
\begin{tabular}{lcccccc}
\toprule
\textbf{Model} & \textbf{Self-B.}$\downarrow$ & \textbf{D-1}$\uparrow$ & \textbf{D-2}$\uparrow$ & \textbf{Gram./100w}$\downarrow$ & \textbf{RR}$\downarrow$ & \textbf{Flesch}$\uparrow$ \\
\midrule
\rowcolor{redcolor} Real      & 0.10 & 0.06 & 0.29 & 5.95 & 0.02 & 27.27 \\
CXR-IRGen & \underline{0.27} & \underline{0.05} & \underline{0.22} & \textbf{2.66} & \textbf{0.00} & \textbf{37.69} \\
ProgEmu   & \textbf{0.16} & \textbf{0.08} & \textbf{0.28} & 4.49 & \underline{0.02} & \underline{34.34} \\
Qwen      & 0.33 & 0.03 & 0.11 & \underline{3.17} & 0.01 & 8.52 \\
\bottomrule
\end{tabular}%
}
\end{table}

\subsection{Deepfake Detection}
\label{sec:detection-results}
We utilize pretrained MedSigLIP~\cite{sellergren2025medgemma} for coherent multimodal deepfake detection. Further, we involve pretrained RAD-DINO~\cite{perezgarcia2024raddino} and CXR-BERT~\cite{boecking2022cxrbert}  as the backbone for image and report deepfake detection pipelines, respectively. We employ a two stage finetuning scheme. Only the classification head is trained during first stage of finetuning. In the second stage, last layer of the backbone is also finetuned along with the classification head. The classification head learning rate is set as $1e-3$ and that for backbone finetuning is $1e-5$. Both stages run for 1,125 steps each with a batch size of 16. Results for all the methods are shown in Table \ref{tab:detection_results}. For human evaluation, accuracy for detecting real vs. fake for different methods were: 75\% for CXR-IRGen, 87.5\% for ProgEMU and 60\% for Qwen. 

\begin{table}[!h]
\centering
\caption{Binary deepfake detection performance.}
\label{tab:detection_results}
\resizebox{\columnwidth}{!}{%
\begin{tabular}{lcccc}
\toprule
\textbf{Model} & \textbf{Accuracy}$\uparrow$ & \textbf{Prec.}$\uparrow$ & \textbf{Rec.}$\uparrow$ & \textbf{F1}$\uparrow$\\
\midrule

\rowcolor{bluecolor} Evaluators (Images) & {92.50} & \textbf{100.00} & {90.00} & \underline{94.70} \\
\rowcolor{bluecolor} RAD-DINO (Images) & \underline{94.67} & {91.25} & \underline{96.35} & {93.34} \\

\rowcolor{greencolor} Evaluators (Reports) & {53.80} & {74.40} & {58.30} & {65.30} \\
\rowcolor{greencolor} CXR-BERT (Reports) & \underline{91.63} & \underline{87.46} & \underline{94.08} & \underline{89.82} \\

MedSigLIP (Image + Rep) & \textbf{99.78} & \underline{98.72} & \textbf{99.68} & \textbf{99.70} \\
\bottomrule
\end{tabular}}
\end{table}

\section{Discussion and Conclusion}
\label{sec:discussion}



Our evaluation shows that current multimodal generation methods can produce individually plausible \ac{CXR}s and radiology reports, but maintaining consistency between the two remains challenging. CXR-IRGen achieves the strongest automated image–report coherence, with matching accuracies ranging from 46.8\% to 67.0\% compared with 78.4\% to 85.0\% for real data. ProgEmu and Qwen show larger coherence gaps, while CXR-IRGen performs strongly on pathology and text fidelity and ProgEmu achieves higher image distributional fidelity in the RAD-DINO feature space. These results indicate that realism of individual modalities does not necessarily translate to clinically consistent image–report pairs. Further, CXR-IRGen's high performance may be attributed to the fact that it generates images with the same labels as the input image and is not a counterfactual image generation model as opposed to ProgEmu and Qwen.

Human evaluation reveals a different pattern from automated assessment. Qwen-generated samples are hardest for human raters to identify as fake (60.0\% detection accuracy vs 75.0\% for CXR-IRGen and 87.5\% for ProgEmu), and achieve the highest human image–report matching accuracy despite the lowest automated matching scores. This discrepancy highlights limitations of relying solely on automated metrics for medical deepfake evaluation. Finally, the MedSigLIP-based multimodal detector achieves 99.78\% accuracy and 99.70\% F1, substantially outperforming unimodal image- and report-based detectors. This result suggests that inconsistencies distributed across the image and report can provide complementary evidence for deepfake detection. Overall, these findings motivate medical deepfake evaluation that jointly considers pathology fidelity, cross-modal consistency, human perception, and multimodal detection.

\bibliographystyle{IEEEbib}
\bibliography{refs}

\end{document}